\documentclass[a4paper,aps,prl,twocolumn,superscriptaddress,nobibnotes,showpacs,]{revtex4-1}
\usepackage[utf8x]{inputenc}
\usepackage[english]{babel}
\usepackage{fontenc}
\usepackage{graphicx}
\usepackage{color}
\usepackage{textcomp}
\usepackage{multirow}
\usepackage{todonotes}
\usepackage{hyperref}
\usepackage{amsmath,amssymb,amsbsy}
\usepackage{siunitx}
\usepackage{orcidlink}

\begin{document}

\title{
Extending the low-$Z$ ``border'' of the $A=100$ region of deformation with precision mass spectrometry of $^{96-98}$Kr
}


\author{D.~Lunney\orcidlink{0000-0002-3227-305X}}
\altaffiliation{
present address:
CNRS International Research Laboratory for 
Nuclear Physics, Nuclear Astrophysics and Accelerator Technologies, TRIUMF, Vancouver, BC, Canada (david.lunney@cnrs.fr)}
\affiliation{Universit\'e Paris-Saclay, IN2P3/CNRS, IJCLab, 91405 Orsay, France}

\author{V.~Manea\orcidlink{0000-0003-2065-5517}}
\altaffiliation{vladimir.manea@ijclab.in2p3.fr}
\affiliation{Universit\'e Paris-Saclay, IN2P3/CNRS, IJCLab, 91405 Orsay, France}
\affiliation{Max-Planck-Institut f\"{u}r Kernphysik, Saupfercheckweg 1, 69117 Heidelberg, Germany}
\affiliation{CERN, 1211 Geneva, Switzerland}

\author{M.~Mougeot\orcidlink{0000-0003-1372-1205}}
\affiliation{Universit\'e Paris-Saclay, IN2P3/CNRS, IJCLab, 91405 Orsay, France}
\affiliation{Max-Planck-Institut f\"{u}r Kernphysik, Saupfercheckweg 1, 69117 Heidelberg, Germany}
\affiliation{CERN, 1211 Geneva, Switzerland}

\author{L.~Nies\orcidlink{0000-0003-2448-3775}}
\affiliation{CERN, 1211 Geneva, Switzerland}
\affiliation{Universit\"{a}t Greifswald, Institut f\"{u}r Physik, 17487 Greifswald, Germany}

\author{N.A.~Althubiti\orcidlink{0000-0003-1513-0409 }}
\affiliation{Physics Department, College of Science, Jouf University, Sakaka, Kingdom of Saudi Arabia}
\affiliation{The University of Manchester, School of Physics and Astronomy, Oxford Road, M13 9PL Manchester, United Kingdom}

\author{D.~Atanasov\orcidlink{0000-0002-0491-4710}}
\altaffiliation{Present Address: Belgian Nuclear Research Centre SCK CEN, Boeretang 200, 2400 Mol, Belgium}
\affiliation{Max-Planck-Institut f\"{u}r Kernphysik, Saupfercheckweg 1, 69117 Heidelberg, Germany}

\author{K.~Blaum\orcidlink{0000-0003-4468-9316}}
\affiliation{Max-Planck-Institut f\"{u}r Kernphysik, Saupfercheckweg 1, 69117 Heidelberg, Germany}


\author{A.~Herlert\orcidlink{0000-0003-1619-0964}}
\affiliation{FAIR GmbH, Planckstra\ss e 1, 64291 Darmstadt, Germany}

\author{W.-J.~Huang\orcidlink{0000-0002-5553-3942}}
\altaffiliation{Present address: Advanced Energy Science and Technology Guangdong Laboratory, 516007, Huizhou, China}
\affiliation{Universit\'e Paris-Saclay, IN2P3/CNRS, IJCLab, 91405 Orsay, France}

\author{J.~Karthein\orcidlink{0000-0002-4306-9708}}
\altaffiliation{Present address: Cyclotron Institute, Department of Physics, Texas A\&M University, TX 77840, USA}
\affiliation{Max-Planck-Institut f\"{u}r Kernphysik, Saupfercheckweg 1, 69117 Heidelberg, Germany}
\affiliation{CERN, 1211 Geneva, Switzerland}

\author{I.~Kulikov\orcidlink{0000-0003-2617-5020}}
\affiliation{GSI Helmholtzzentrum f\"{u}r Schwerionenforschung GmbH, Planckstra\ss e 1, 64291 Darmstadt, Germany}

\author{Yu.~A.~Litvinov\orcidlink{0000-0002-7043-4993}}
\affiliation{GSI Helmholtzzentrum f\"{u}r Schwerionenforschung GmbH, Planckstra\ss e 1, 64291 Darmstadt, Germany}
\affiliation{Institut f\"ur Kernphysik, Universit\"at zu K\"oln, Z{\"u}lpicher Str. 77, D-50937 K\"oln, Germany}

\author{D. Neidherr\orcidlink{0000-0002-7332-9144}}
\affiliation{GSI Helmholtzzentrum f\"{u}r Schwerionenforschung GmbH, Planckstra\ss e 1, 64291 Darmstadt, Germany}

\author{T.R. Rodr\'{\i}guez\orcidlink{0000-0002-3516-8239}}
\altaffiliation{Present address: Departamento de F\'isica At\'omica, Molecular y Nuclear, Universidad de Sevilla, E-41012 Sevilla (Spain)}
\affiliation{Departamento de Estructura de la Materia, Física Térmica y Electrónica and IPARCOS, Universidad Complutense de Madrid, E-28040 Madrid, Spain}

\author{M. Rosenbusch\orcidlink{0000-0002-5199-1073}}
\altaffiliation{Present Address: RIKEN Nishina Center for Accelerator-Based Science, Wako, Saitama 351-0198, Japan}
\affiliation{Universit\"{a}t Greifswald, Institut f\"{u}r Physik, 17487 Greifswald, Germany}


\author{L. Schweikhard\orcidlink{0009-0002-8272-0388}}
\affiliation{Universit\"{a}t Greifswald, Institut f\"{u}r Physik, 17487 Greifswald, Germany}

\author{T. Steinsberger\orcidlink{0000-0002-2388-8706}}
\affiliation{Max-Planck-Institut f\"{u}r Kernphysik, Saupfercheckweg 1, 69117 Heidelberg, Germany}

\author{A. Welker\orcidlink{0000-0003-4435-0208}}
\affiliation{CERN, 1211 Geneva, Switzerland}
\affiliation{Technische Universit\"{a}t Dresden, 01069 Dresden, Germany}

\author{F.~Wienholtz\orcidlink{0000-0002-4367-7420}}
\altaffiliation{Present Address: Institut f\"ur Kernphysik, Technische Universit\"at Darmstadt, 64289 Darmstadt, Germany}
\affiliation{Universit\"{a}t Greifswald, Institut f\"{u}r Physik, 17487 Greifswald, Germany}

\author{R.N.~Wolf\orcidlink{0000-0003-0748-7387}}
\affiliation{Max-Planck-Institut f\"{u}r Kernphysik, Saupfercheckweg 1, 69117 Heidelberg, Germany}
\affiliation{ARC Centre of Excellence for Engineered Quantum Systems,The University of Sydney, NSW 2006, Australia}

\date{\today}

\begin{abstract}
The onset of collective nuclear behavior in the ${N=60}$, ${A\sim100}$, region is examined through high-precision mass measurements of $^{96-98}$Kr, performed with the ISOLTRAP mass spectrometer at ISOLDE, CERN.
Our results for $^{96-97}$Kr agree with previous measurements, with our new $^{97}$Kr Penning-trap mass value three times more precise.
The mass value of $^{98}$Kr is measured for the first time.  The new mass surface, together with comparisons to beyond-mean-field theoretical predictions, suggests that collectivity persists for the ${Z=36}$ isotopes, blurring the apparent ``low-$Z$ boundary'' of this deformed region. 
\end{abstract}


\maketitle

\section{Introduction}


Neutron-rich ${A=100}$ nuclides around ${Z=40}$ exhibit one of the most dramatic nuclear-structure changes on the nuclear chart, manifested by a pronounced discontinuity at neutron number ${N=60}$ in the trends of several nuclear properties, including binding energies, mean-square charge radii, and excitation spectra.
From the first experimental evidence in this region, obtained through the spectroscopy of $^{252}$Cf fission fragments~\cite{JOHANSSON}, the change in structure was linked to the phenomenon of nuclear deformation.
The emerging picture is thus one of a sudden transition from spherical or weakly deformed below ${N = 60}$, to strongly deformed above. 

This evolution has stimulated many theoretical developments, being a well-known example of the dramatic effect the proton-neutron interaction can have on driving structural changes.
In the case of zirconium isotopes, in which the discontinuity of nuclear properties at ${N = 60}$ is the most pronounced, early calculations with the configuration interaction method~\cite{Federman79} have emphasized the role of the strong attractive interaction between neutrons filling the $g_{7/2}$ orbital from ${N = 60}$ on and protons excited to the $g_{9/2}$ orbitals above the ${Z = 40}$ gap.
This interaction leads to a large gain in binding energy, which overcompensates the energy cost of exciting the protons. The resulting state, which is considerably deformed, thus becomes the ground state through this binding energy gain. The echoes of this work can be found in more recent theoretical studies such as \cite{Togashi2016}, where the tensor interaction, in addition to the central part of the proton-neutron interaction, was shown to lead to so-called Type II shell evolution. Other works \cite{Urban} have also underlined the role of the extruder neutron $g_{9/2}$ orbital (in addition to that of the intruder $g_{9/2}$ proton orbital) in driving the accelerated occupation of $g_{7/2}$ by valence neutrons. 

The early theoretical work concerning the $N = 60$ shape transition had already predicted that removing a pair of protons from ${Z = 40}$ would empty the $p_{1/2}$ proton shell, significantly increasing the cost of exciting protons to $g_{9/2}$ from the even deeper $p_{3/2}$ orbital.
It was therefore expected that with the reduction of proton number, 
the effects leading to the dramatic ${N = 60}$ shape change would be less prominent for strontium ($Z=38$) and krypton ($Z=36$).  
Nevertheless, as discussed in~\cite{Federman84}, a reduction of the $p_{3/2}$ - $p_{1/2}$ proton gap with increasing neutron number causes strontium to also experience this deformation-driving mechanism.  Therefore, krypton becomes the first even-$Z$ isotopic chain where the mechanism might break down. 

Over the last decades, increasingly precise measurements of the masses of the neutron-rich, ${A=100}$ nuclides have been accomplished.
The advent of Penning traps allowed refining the trends of the mass surface in the region, starting with measurements of the refractory zirconium isotopes from the IGISOL facility~\cite{Rinta}, later expanded to heavier (technetium, ruthenium, rhodium and palladium) \cite{Hager} and lighter (strontium, rubidium) \cite{Rahaman} chains. 
These trends were extended significantly beyond ${N = 60}$ in the strontium and rubidium isotopic chains with mass measurements performed using ISOLTRAP~\cite{Manea, deRoubin} and TITAN~\cite{Mukul}, and most recently in yttrium, zirconium, niobium, and molybdenum with JYFLTRAP~\cite{HUKKANEN2024138916}. 

The report of a very low-energy $2^{+}$ state in $^{96}$Kr \cite{Marginean2009} spurred the continuation of mass measurements along the krypton chain with ISOLTRAP \cite{DELAHAYE2004604}, reaching 
$^{97}$Kr \cite{Naimi}, crossing the ${N = 60}$ isotonic line. However, no discontinuity in the two-neutron separation energies was found, in accordance with earlier laser spectroscopy results from COLLAPS~\cite{Keim}, which had found no discontinuity in the charge radii.
Coulomb-excitation measurements at REX-ISOLDE confirmed this picture, finding no significant lowering in the energy of the first excited 2$^+$ state in $^{96}$Kr after all~\cite{Albers}.
More recently, the excitation energies of the first 2$^+$ states in krypton were measured up to $A = 100$ at RIKEN-RIBF~\cite{Flavigny}.
Detailed spectroscopy measurements were performed in $^{96}$Kr by Dudouet et al.~\cite{Dudouet} using the AGATA array at GANIL and in $^{96,98}$Sr with REX-ISOLDE~\cite{clement}. All of these results have illustrated the markedly different evolution of excited states in the krypton isotopic chain with respect to the higher-$Z$ ones, including a lowering of the ratio between the energies of the first $4^+$ and $2^+$ states in $^{96}$Kr, at apparent odds with a picture of increasing deformation.
 
The mean-field or density-functional theoretical approaches offer a natural framework to describe the phenomenon of nuclear deformation by allowing the nuclear many-body wavefunction to break rotational symmetry.
It is thus possible to identify the so-called intrinsic nuclear shapes as variational minima of the nuclear binding energy as a function of the deformation parameters~\cite{Ring00}.
Subsequently, symmetry restoration and other beyond-mean-field many-body methods allow recovering the physical ground and excited states with the appropriate quantum numbers and offer the possibility to link the effect of the intrinsic nuclear shapes to experimental data~\cite{Bender06,Delaroche10}. 

For the ${A \approx 100}$ nuclei, in the last decades, there have been different self-consistent mean-field studies either dedicated to the ${N = 60}$ shape transition, or producing global tables that also cover the region in question.
Systematically, these approaches have predicted that the nuclides around ${N = 60}$ possess an oblate intrinsic shape and two prolate shapes, one less deformed which is present also below ${N = 60}$ and one very deformed which suddenly emerges around ${N = 60}$~\cite{Rodriguez-Guzman10,Rodriguez-Guzman10b,UNEDF0}.
The ordering in energy between the oblate and prolate shapes and the accuracy with which these models describe the trends of ground-state observables can vary significantly and they are, as discussed in~\cite{Manea}, within the same model, sensitive to details of the nuclear interaction such as the pairing strength.
Nevertheless, the emerging picture is that at ${N = 60}$ a transition from the oblate to the prolate structure takes place in the ground state, which corresponds
to a sudden increase in mean-square charge radii and a flattening or increase in two-neutron separation energies. 

In this picture, it becomes apparent that the same shapes and competition around ${N = 60}$ are also present in the krypton isotopic chain.  Thus it remains an open question as to whether krypton would exhibit a smooth shape evolution, or a shape transition would occur beyond ${N = 60}$.
Another question is if the mean-field picture would be altered by triaxiality, or whether symmetry restoration and beyond-mean-field shape fluctuations would significantly impact the trends of nuclear observables.

In this work, we extend the knowledge of ground-state binding energies in the krypton isotopic chain, presenting the first mass value for $^{98}$Kr, as well as an improvement of the masses of $^{96}$Kr and $^{97}$Kr.
We interpret the resulting trends of two-neutron separation energies with respect to both density-functional~\cite{UNEDF0} and beyond-mean-field~\cite{Rodriguez2014} calculations.

\section{Experiment}

\begin{figure*}
\centering
\includegraphics[width=1\textwidth]{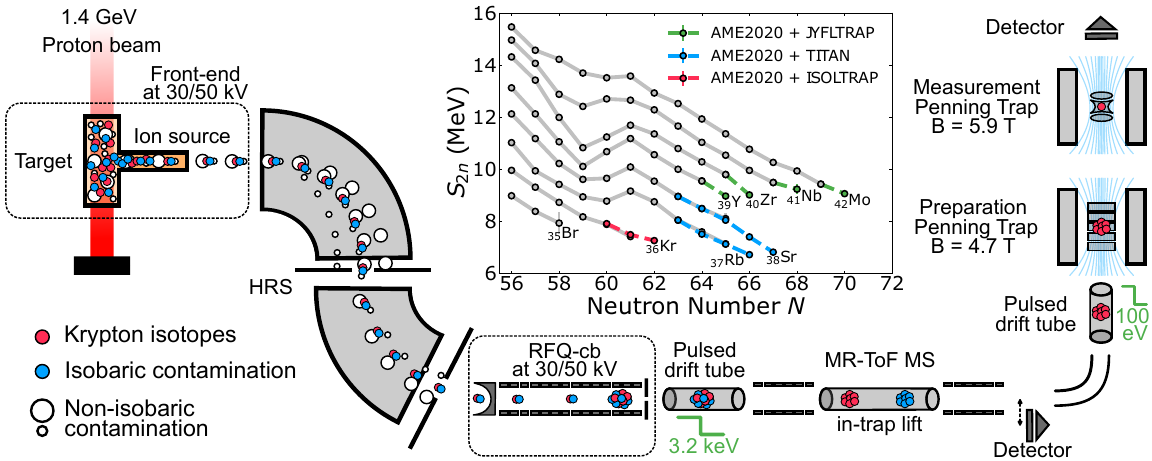}
\caption{Schematic representation of ISOLDE and the ISOLTRAP on-line mass spectrometer. The typical kinetic energy of the ions at various stages of the ISOLTRAP apparatus is shown in green (for details, see~\cite{Mukherjee-EurPhysJA,Kreim-NuclInstrumMethodsB.317.492}).
The insert shows the two-neutron separation energies (from the AME2020 \cite{AME2020}) versus neutron number ($N$) for the $A=100$, $N=60$ region with the new masses presented here highlighted in red.  Also shown are more recent mass measurements for neutron-rich Rb, Sr, Y, Zr, Nb, and Mo isotopes from~\cite{Mukul, HUKKANEN2024138916}.}
\label{isoltrap_sketch}
\end{figure*}

Our measurements were performed at the radioactive ion-beam facility ISOLDE at CERN~\cite{ISOLDE_2017} in July 2015 and August 2017.
The radioisotopes of interest were produced by impinging a primary beam of $\SI{1.4}{\giga\electronvolt}$ protons delivered by CERN's PS Booster on a thick UC$_{x}$ target, shown schematically on the top left in Fig.~\ref{isoltrap_sketch}.
A Versatile Arc Discharge Ion Source (VADIS VD7) was used for ionizing reaction products that diffused out of the heated target container~\cite{doi:10.1063/1.3271245}.
To inhibit the effusion of the less volatile species into the source, a water-cooled tantalum transfer line was employed.
The obtained flux of ions was accelerated to a kinetic energy of 30/50 keV in 2015/2017, respectively.
The mass-over-charge $m/q$ selection of the isobars of interest was performed using the ISOLDE High-Resolution (magnetic-dipole) Separator (HRS), depicted as the cut circles downstream of the ion source in Fig.~\ref{isoltrap_sketch}. 

A schematic representation of the ISOLTRAP mass-spectrometer~\cite{Mukherjee-EurPhysJA,Kreim-NuclInstrumMethodsB.317.492} is shown on the right-hand side of Fig.~\ref{isoltrap_sketch}. 
The quasi-continuous ion beam from HRS was 
transported to the ISOLTRAP spectrometer, where it was electrostatically decelerated to less than $\SI{100}{\electronvolt}$ and injected into a linear radio-frequency cooler-buncher trap (RFQ-cb)~\cite{Herfurth2001254}, where the emittance was reduced in a few milliseconds through collisions with helium buffer gas.
However, as a noble gas, 
krypton ions are prone to charge-exchange reactions with the neutral impurities contained in the helium gas of the RFQ-cb~\cite{DELAHAYE2004604}.
To mitigate losses of the short-lived krypton ions of interest, the helium gas fed to the RFQ-cb was purified using a liquid nitrogen-filled cold trap, and the charge-exchange half-life of singly charged $^{91}$Kr was determined to be $~\SI{62}{\milli\second}$.

Subsequently, ions were extracted in short bunches after $\SI{10}{\milli\second}$ of cooling and bunching time, followed by deceleration by a pulsed drift tube to a kinetic energy of $\approx\SI{3.2}{\kilo\electronvolt}$, and injected into a Multi-Reflection Time-of-Flight Mass Separator (MR-ToF MS)~\cite{WOLF201282,WOLF2013123}. 
In all cases, the short-lived radioactive species were unambiguously identified by turning the proton beam off and observing the effect of on the recorded time-of-flight spectra. Other identification tests were performed for $^{97,98}$Kr, as detailed below.

The mass measurements were performed using two methods, namely the MR-ToF time-of-flight and Penning-trap time-of-flight ion-cyclotron resonance (ToF-ICR) mass spectrometry techniques.
In the case of low yield and/or short half-life only the MR-ToF MS was used.
The fundamental relationship of time-of-flight mass spectrometry
\begin{equation}
\label{eqn:tof}
t_{x} = a \sqrt{\frac{m_{ion,x}}{q_{x}}} + b,
\end{equation}
links an ion's mass-over-charge ratio $\frac{m_{\textit{ion,x}}}{q_{x}}$ to its time-of-flight $t_{x}$~\cite{Guilhaus1995}.
In Equation~\ref{eqn:tof}, $a$ and $b$ are calibration parameters that can be determined by measuring the flight times $t_{1,2}$ of two reference ions with well-known masses $m_{1,2}$ and charges $q_{1,2}$, provided they travel the same distance as the ion of interest, i.e. the same number of revolutions inside the device.
The mass of an ion is then calculated from the relation~\cite{Wienholtz-Nature.498.346}:
\begin{align} 
\sqrt{\frac{m_{ion,x}}{q_{x}}} & = C_{\textit{TOF}} \left( \sqrt{\frac{m_{ion,1}}{q_{1}}}-\sqrt{\frac{m_{ion,2}}{q_{2}}} \right) \nonumber \\
 & + \frac{1}{2} \left( \sqrt{\frac{m_{ion,1}}{q_{1}}}+\sqrt{\frac{m_{ion,2}}{q_{2}}} \right),
\end{align}
with :
\begin{equation}
C_{TOF} = \frac{2t_{x}-t_{1}-t_{2}}{2(t_{1}-t_{2})}.
\end{equation}

The second technique used in this work is the well-established Time-of-Flight Ion-Cyclotron-Resonance of ions in a Penning trap~\cite{Koenig-IntJMassSpectrom.142.95}, which determines an ion's mass $m_{\textit{ion,x}}$ from the ion's free cyclotron frequency according to:
\begin{equation}
\label{eqn:toficr}
\nu_{c,x} = \frac{q_{x} B}{2 \pi m_{\textit{ion,x}}},
\end{equation}
where $q_{x}$ is the ion's charge and $B$ is the strength of the Penning-trap magnetic field.

When the yield and half-life were sufficient, the MR-ToF MS was used as a mass filter to select the species of interest by optimizing the timing and length of the extraction pulse from the MR-ToF MS~\cite{WIENHOLTZ2017285} to only extract the ion of interest towards the Penning traps.
There, further purification was achieved using a well-established, mass-selective, buffer-gas cooling technique~\cite{SAVARD1991247}, transported through a 90~degree bend to ISOLTRAP's vertical section and captured in the preparation Penning trap~\cite{RAIMBAULTHARTMANN1997378}. 
The obtained ion bunch was finally transported to ISOLTRAP's hyperbolic, precision Penning trap, where the free cyclotron frequency measurement was performed.
To avoid systematic shifts in cyclotron frequencies due to space charge inside the trap, only one ion on average was allowed to remain in the precision trap during the measurement.

In Equation~\ref{eqn:toficr}, the calibration of the magnetic field is performed by measuring the cyclotron frequency $\nu_{\textit{c,ref}}$ of a reference species of well-known mass $m_{\textit{ion,ref}}$ shortly before and shortly after the measurement of the species of interest. The cyclotron frequency of the reference species is then linearly interpolated to the time at which the measurement of the ion of interest was performed.
From the experimentally measured cyclotron-frequency ratio
\begin{equation}
R = \frac{\nu_{c,\textit{ref}}}{\nu_{c,x}} = \frac{m_{ion,x}\cdot q_{\textit{ref}}}{m_{ion,\textit{ref}}\cdot q_\textit{x}},
\end{equation}
the atomic mass $m_\textit{a,x}$ of the species of interest is calculated according to the relation from~\cite{Gallant2012}: 
\begin{equation}
m_{\textit{a,x}} = \frac{q_\textit{x}}{q_\textit{ref}}R(m_{\textit{a,ref}} - q_\textit{ref}m_{e}+B_\textit{e,ref}) + q_\textit{x}m_{e} - B_\textit{e,x},
\label{eq:freq_ratio_R}
\end{equation}
where $m_{e}$ is the electron mass~\cite{Sturm2014}, $m_\textit{a,ref}$ is the atomic mass of the reference species, and $B_\textit{e,ref}$ and $B_\textit{e,x}$ are the electron binding energies of the missing electrons of the measured reference ion and ion of interest, respectively. 
In most cases only singly-charged ions are measured in the Penning trap leading to $q_x = q_\textit{ref} = 1$.
Furthermore, as the electron binding energy of a singly-charged ion is in the order of a few eV and the measurement uncertainty of the online Penning trap spectrometry is in the order of a few keV, $B_\textit{e,ref}$ and $B_\textit{e,x}$ can be neglected, which simplifies formula \eqref{eq:freq_ratio_R} to
\begin{align}
m_{\textit{a,x}} = R(m_{\textit{a,ref}} - m_{e}) + m_{e}.
\end{align}

\begin{table*}[!ht]
\centering
 \caption{Final frequency ratios ($R$), time-of-flight ratios ($C_{ToF}$) and mass excesses of the krypton isotopes measured in this work. Values of the mass excesses from the Atomic-Mass Evaluation 2020 (AME2020)~\cite{AME2020} are given for comparison.  Note the \char"0023~symbol indicates a value derived from systematics. Experimental half-lives are taken from the NUBASE2020 evaluation~\cite{Kondev2021}.} 
\begin{ruledtabular}
  \begin{center}
    \begin{tabular}{c c c c c c}
    & & & & \multicolumn{2}{c}{\textbf{Mass Excess (keV)}} \\ \cline{5-6} 
    \textbf{Species} & \textbf{Half-life} & \textbf{Reference Ions} & \textbf{ ratio $R$ or $C_{ToF}$} & \textbf{This work} & \textbf{AME2020} \\ \hline
    $^{96}$Kr$^{2+}$ & 80(8) ms & $^{32}$S$^{16}$O$^{+}$/$^{85}$Rb$^{+}$ & $C_{ToF}$ = $\SI{0.499857808\pm 0.000000161}{}$  & $-\SI{53106\pm 10}{}$ & $\SI{-53082 \pm 19}{}$\\  \cline{1-6}
    \multirow{2}{*}{$^{97}$Kr$^{+}$} &  \multirow{2}{*}{62.2(3.2) ms}  & $^{39}$K$^{+}$ & R = $\SI{2.488208221(0.000000989)}{}$  & $-\SI{47508\pm 36}{}$ & \multirow{2}{*}{$-\SI{47420\pm 130}{}$}\\
        &  &   $^{97}$Mo$^{+}$/$^{85}$Rb$^{+}$ & $C_{ToF}$ = $\SI{0.50347338\pm 0.00000603}{}$ & $-\SI{47452\pm 71}{}$ &\\ \hline
    $^{98}$Kr$^{+}$ & 42.8(3.6) ms & $^{98}$Mo$^{+}$/$^{85}$Rb$^{+}$ & $C_{ToF}$ = $\SI{0.50349909\pm 0.00000405}{}$
    & $\SI{-44249\pm 52}{}$ & $-\SI{44120}{}$\#(300\#)\\
        \end{tabular}
   \end{center}
  \end{ruledtabular}
   \label{Kr_results}
\end{table*}

\subsection{Mass measurement of $^{96}$Kr}

\begin{figure}
  \centering 
\includegraphics[width=\columnwidth]{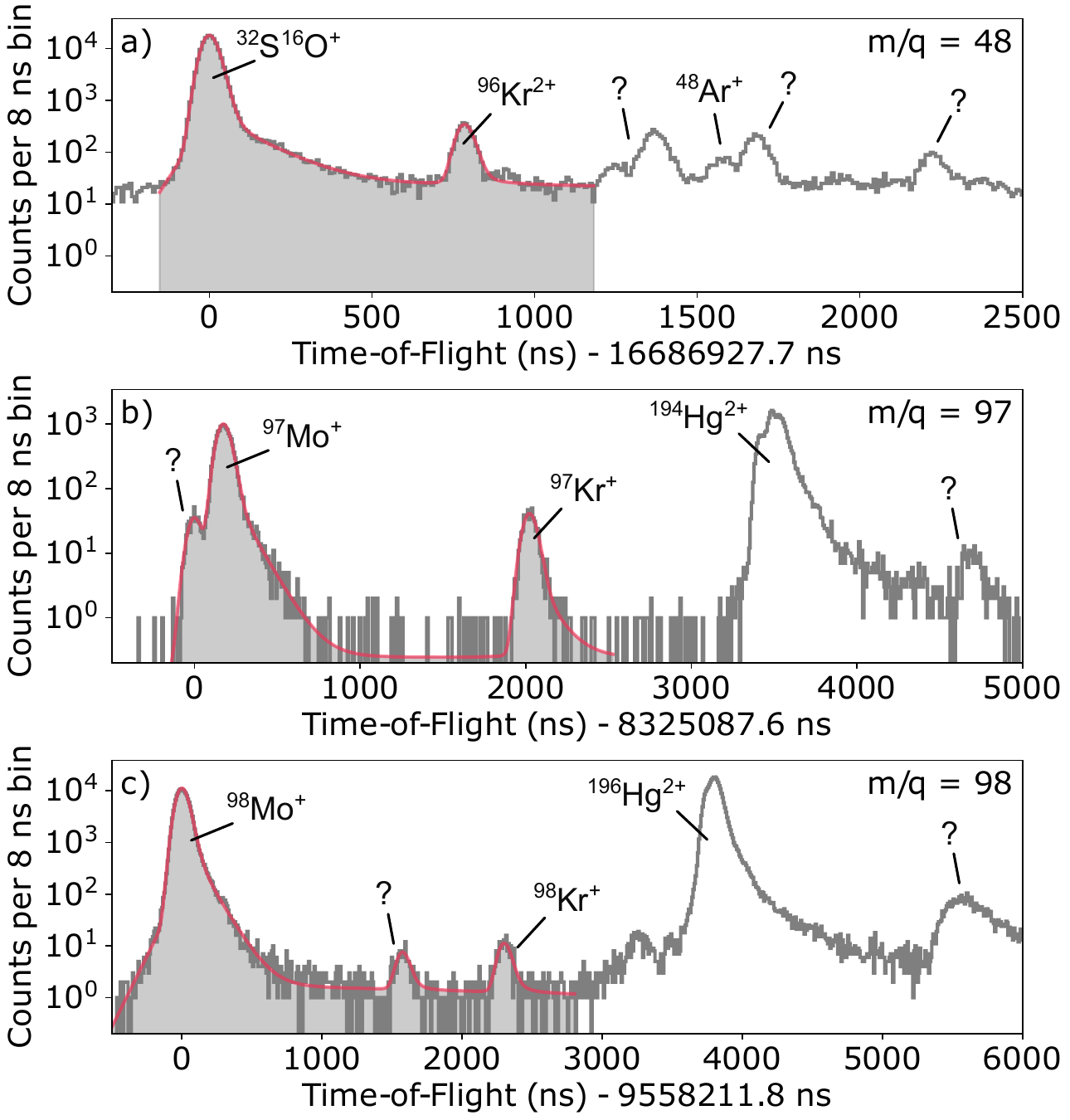}
 \caption{Time-of-flight spectra for the different MR-ToF MS measurements. a) $m/q=48$ spectrum recorded after 1000 revolutions, b) after 300 for $m/q=97$, c) and after 400 revolutions for $m/q=98$. 
 The solid red lines represent the hyperEMG fits to the data while the shaded area indicates the fit range.
 Identified species are indicated, unidentified species are marked with a question mark.}
\label{MRToF_spectra}
\end{figure}

The atomic mass value of $^{96}$Kr reported here was measured in 2017 using the MR-ToF MS technique.
The neutron-rich krypton isotopes of interest were measured as doubly-charged ions as a side product of a dedicated $^{48}$Ar$^{+}$ experiment~\cite{PhysRevC.102.014301}.
Figure~\ref{MRToF_spectra}a) shows a typical ${m/q=48}$ time-of-flight spectrum obtained after 1000 revolutions inside the MR-ToF MS.
The dominant contaminant species in the radioactive ion beam was found to be $^{32}$S$^{16}$O$^{+}$ and was used as an online reference species.
ISOLTRAP's offline surface-ionization source was used to deliver $^{85}$Rb$^{+}$ reference ions.
In total, about 2600 $^{96}$Kr$^{2+}$ ions were recorded.

Exponentially-Modified Gaussian (EMG) probability-density functions consisting of several different exponential components were used to extract the time-of-flight of the ions of interest by means of the binned maximum likelihood method, implemented using CERN's RooFit MINUIT optimizer~\cite{James:2296388}. 
The resulting~\textit{hyperEMG} fit model by Purushothaman \textit{et al}.~\cite{2017_Purushothaman_hyperEMG}, describes an asymmetric Gaussian distribution with a number of exponential tails on both sides of the distribution.
In the present case, the spectra were fitted with models including one negative tail (towards faster times-of-flight) and up to three positive tails (towards slower times-of-flight) to account for the characteristic tailing of the MR-ToF MS data and the almost uniform background observed in some of the spectra.

To quantify space-charge effects in this data set, i.e. the Coulomb interaction of ions stored in the device that can lead to time-of-flight shifts~\cite{doi:10.1063/1.4796061, Maier2023spacecharge}, the sum data file was separated into spectra with similar numbers of ions in the corresponding measurement cycle.
By varying the number of charges allowed inside the trap, their spectra were fit and the stability of the resulting mass values was examined.
As a lower limit for each sub-set, about 10\% of the total krypton counts of the sum data file was needed to guarantee a stable fit.
The resulting uncertainty of the weighted average of the extracted mass excess, determined to be $\SI{7}{\kilo\electronvolt}$, was then added in quadrature to the statistical uncertainty of the fit to the sum data.
This procedure resulted in a mass excess of $-\SI{53106\pm 10}{\kilo\electronvolt}$.


Two measurements contribute to the Atomic Mass Evaluation (AME2020) $^{96}$Kr mass value~\cite{AME2020}.
The first, performed at ISOLTRAP in 2010~\cite{Naimi} using the ToF-ICR technique, contributes 88\% to the AME2020 tabulated value.
A measurement from 2020, performed with the {TITAN} MR-ToF MS at TRIUMF~\cite{TITAN_KR}, accounts for the remaining 12\%.
The $^{96}$Kr mass excess reported here agrees within one standard deviation with these results. As its uncertainty is two times better than the 2010 ISOLTRAP Penning-trap measurement, it now accounts for 81\% with the 2010 ISOLTRAP result providing the remaining 19\%.   

\subsection{Mass measurement of $^{97}$Kr}

\begin{figure}
  \centering 
\includegraphics[scale=0.4]{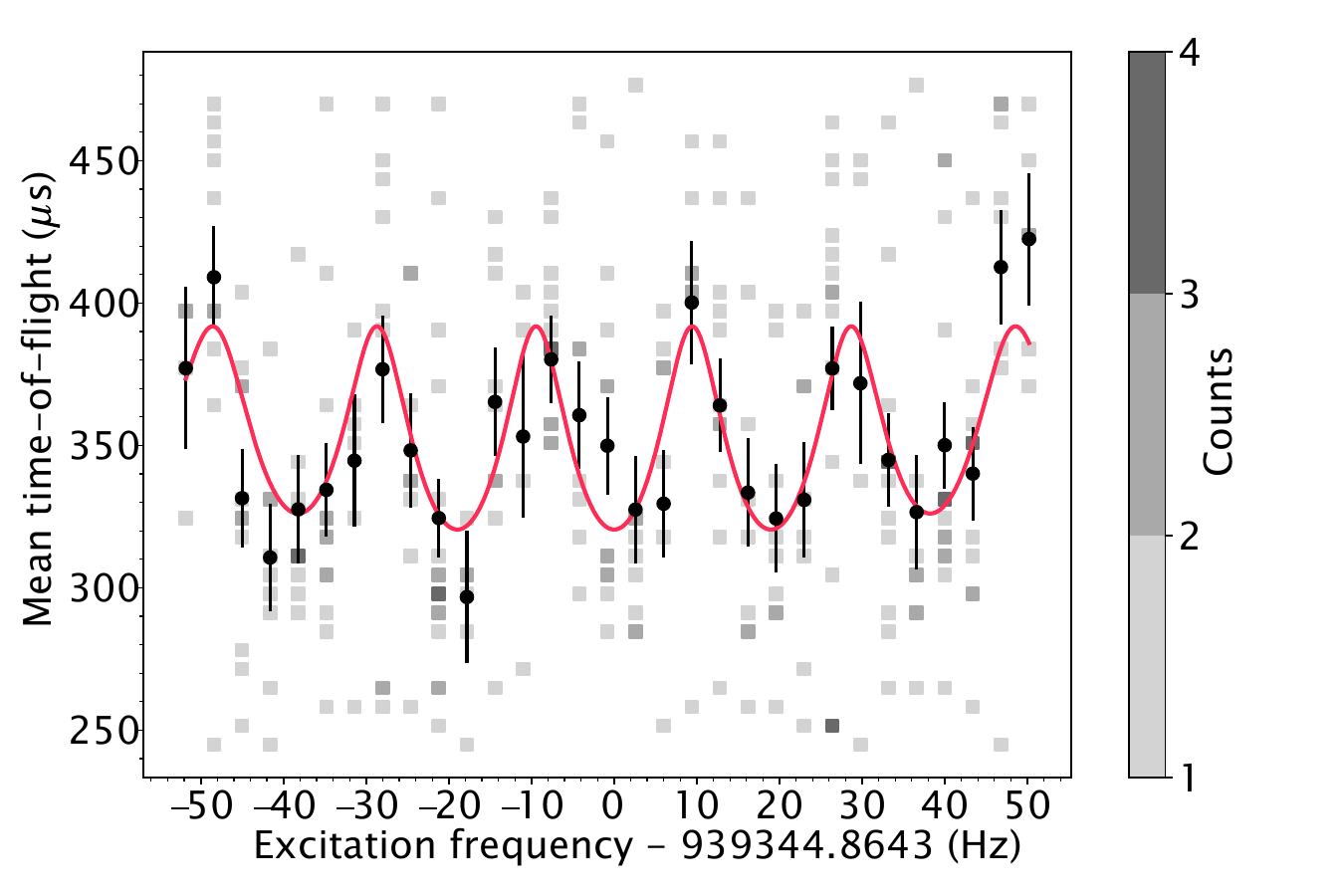}
 \caption{A typical ToF-ICR resonance of $^{97}$Kr$^{+}$ using the Ramsey-type excitation scheme ($T_{on}^{RF}-T_{off}^{RF}-T_{on}^{RF}$ = 10 ms-40 ms -10 ms)~\cite{George-IntJMassSpectrom.264.110,PhysRevLett.98.162501}. The grey-scale map represents the ion events recorded in each bin. The mean and standard deviation of the time-of-flight distribution recorded in each frequency bin are shown as black circles while the red line shows the result of the least-squares adjustment of these data points to the theoretical line shape.}
 \label{97Kr_ramsey}
\end{figure}

A ToF-ICR spectrum of $^{97}$Kr$^{+}$ recorded during the 2015 run using the Ramsey-type excitation scheme ($T_{on}^{RF}-T_{off}^{RF}-T_{on}^{RF}$ = 10~ms~-~40~ms~-~10~ms) \cite{George-IntJMassSpectrom.264.110,PhysRevLett.98.162501} is shown in Fig.~\ref{97Kr_ramsey}.
Additional resonances using a single (50~ms) excitation were also recorded.
In all cases, $^{39}$K$^+$ ions were used as reference masses.
The established analysis protocol \cite{kellerbauer2003} was used to include a mass-dependent systematic error and all spectra were recorded with less than three ions in the trap to avoid shifts due to space-charge.

The frequency ratio $R$ and derived mass excess are listed in Table~\ref{Kr_results}.
It is noteworthy that this result agrees with the mass determined with ISOLTRAP by Naimi \textit{et al}.~\cite{Naimi}.
In that work, charge-exchange losses prevented bringing the $^{97}$Kr ions into the measurement Penning trap so an improvised measurement was made using so-called ``cooling'' scans in the gas-filled preparation Penning trap, involving an empirical fit to the data using a double Wood-Saxon function, described in~\cite{doublewood}.  

During the 2017 run, 
the spectrum shown in Figure~\ref{MRToF_spectra}b) was recorded for $m/q=97$ at 300 revolutions, and about 600 $^{97}$Kr$^{+}$ ions were identified. The opening of the ISOLDE beam gate was synchronized to the impact of the proton pulse on the ISOLDE target. Determining the $^{97}$Kr$^+$ count rate after the MR-TOF MS for different delays between the proton impact and the opening of the beam gate allowed recording a qualitative release curve of $^{97}$Kr from the target. A significant drop of the count rate was observed for delays larger than 100 ms, consistent with the short half-life of $^{97}$Kr. We note that the release curve measured in an identical fashion for $^{91}$Kr$^+$ showed a rather flat behavior of the count rate up to 400 ms delay to the proton pulse.
The abundant and stable $^{97}$Mo$^{+}$ ion, contamination from the ISOL target unit, was used as an online reference, in addition to $^{85}$Rb$^{+}$ ions from the offline ion source.
The strongest radioactive contamination was $^{194}$Hg$^{2+}$, which saturated the data acquisition system and appears to have several peaks in its ToF spectrum. Due to this, only a sub-range of the ToF spectra was fitted.

Using the same analysis procedure described above, we derive a mass excess value of $-\SI{47452\pm 71}{\kilo\electronvolt}$, including a systematic uncertainty of $\SI{14}{\kilo\electronvolt}$ added in quadrature to account for space charge dependent time-of-flight shifts. 
The $C_{tof}$ and derived mass values are listed in Table~\ref{Kr_results}, likewise in excellent agreement with the Ramsey and double-Wood-Saxon values.
Combining the two new measurements gives a weighted mean of $\SI{-47497\pm 32}{\kilo\electronvolt}$, which accounts for 94\% of the new adjusted value.  The remaining 6\% are from the 2010 ISOLTRAP cooler-trap measurement \cite{Naimi}.
This is the first isotope for which ISOLTRAP has measured its mass using three different ion traps and methods.  

\subsection{Mass measurement of $^{98}$Kr}

Due to the lower production yields and half-life, the mass measurement of $^{98}$Kr was performed only using the MR-ToF MS.
Data was taken in 2015 for an ion beam with $m/q=98$ at 350 revolutions and in 2017 at 350 and 400 revolutions.
A total of about 
1000 $^{98}$Kr$^{+}$ ions were recorded during the three different measurements. 
An example spectrum for 400 revolutions data is shown in Fig.~\ref{MRToF_spectra}c).
The abundant (stable) $^{98}$Mo$^+$ was used as a reference mass while radiogenic, doubly charged mercury $^{196}$Hg$^{2+}$ was observed as well.
Two other contaminants could not be identified.


To verify the $^{98}$Kr signal in the spectrum at the expected time of flight, different tests were performed. Firstly, as in the case of $^{97}$Kr, the opening of the beam gate was synchronized to the impact of the proton pulse on the target and a variable delay was included. The behavior of the count rate while increasing the delay was consistent with the isotope half-life. Secondly, for a given beam-gate duration and minimal delay to the proton pulse, the count rate was determined for different holding times of the ions in the ISOLTRAP buncher (5, 100 and 200 ms). The drop in the ion count rate with increased holding time is consistent with the combined charge-exchange and decay half-life of $^{98}$Kr. Finally, the yield drop between $^{96}$Kr-$^{97}$Kr (a factor of 15) and$^{97}$Kr-$^{98}$Kr (a factor of 17) is consistent with what is expected for these increasingly neutron-rich radiogenic isotopes. All these precautions give confidence that the measured mass corresponds to $^{98}$Kr.

The stable $^{98}$Mo ion was chosen as on-line reference due its well known mass together with $^{85}$Rb supplied by the offline ion source.
The $C_{tof}$ and atomic mass were calculated and are listed in Table~\ref{Kr_results}.
To account for systematic space-charge effects, different ion loads were fitted, and the resulting uncertainty was added in quadrature to the fit uncertainty, overall yielding a mass excess of $\SI{-44249\pm 52}{\kilo\electronvolt}$.
This is the first experimental mass determination for $^{98}$Kr and deviates by $\SI{100}{\kilo\electronvolt}$ from the extrapolated AME2020 value~\cite{AME2020}, but well within the derived $\SI{300}{\kilo\electronvolt}$ uncertainty.

\begin{figure*}
\centering 
\includegraphics[width = 0.75\linewidth]{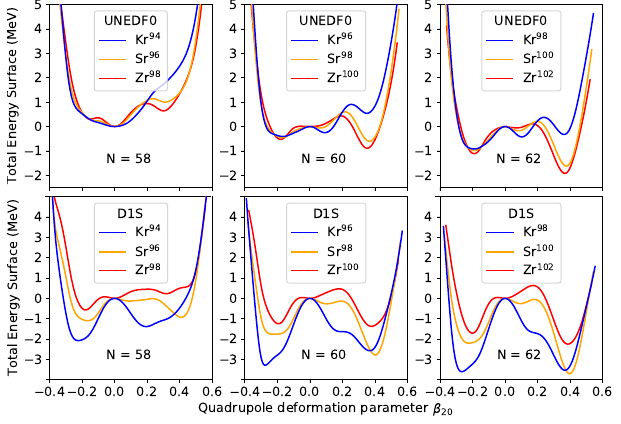}
\caption{Upper row: Total energy as a function of quadrupole deformation $\beta_{20}$ obtained in a calculation with the UNEDF0 functional for the krypton ($Z=36$), strontium ($Z=38$) and zirconium ($Z=40$) isotopes with $N = 58$ (left panel), $N = 60$ (center panel) and $N = 62$ (right panel). The total energy of the spherical solution is subtracted for each case.  Negative values of $\beta_{20}$ correspond to oblate shapes while positive values are prolate. Lower row: same as above but with the Gogny D1S interlaction.
These calculations were performed using the HFBTHO code version 3.00 \cite{HFBTHO}.
}
\label{fig:DFT}
\end{figure*}

\begin{figure*}
  \centering 
\includegraphics[width = 1\linewidth]{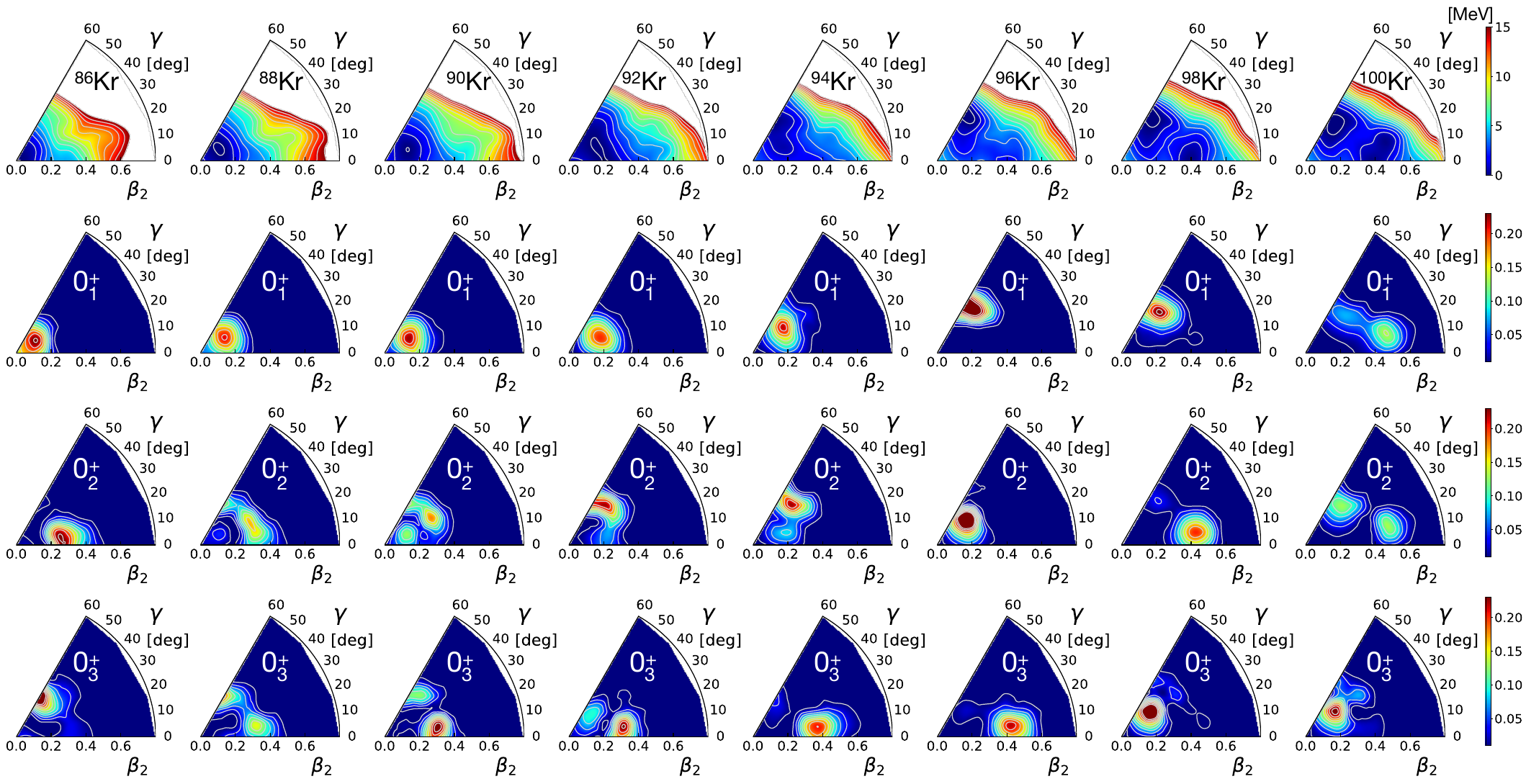}
 \caption{(top row) Total energy surfaces and (bottom rows) collective wavefunctions for the first three excited $0^{+}$ states of $^{86-100}$Kr,
 calculated using the beyond-mean-field SCCM method described in \cite{Rodriguez2014}.  
 }
\label{TES-SCCM}
\end{figure*}


\section{Discussion}

The new masses were used to calculate two-neutron separation energies:  $S_{2n}=BE(Z,N) - BE(Z,N-2)$, where $BE$ is the binding energy (determined from the measured masses).  The krypton $S_{2n}$ values are plotted versus neutron number in the top panel of Fig.~\ref{isoltrap_sketch} with the new masses highlighted in red.  The neighboring isotopic chains (with mass values from \cite{AME2020} and more recent works \cite{Mukul,HUKKANEN2024138916}) are also shown for comparison. The previous ISOLTRAP measurement of $^{97}$Kr \cite{Naimi} is confirmed and with it the trend indicating krypton as the boundary of the shape transition phenomenon at $N = 60$. The additional data point at ${N=62}$ addresses the question of whether some sensitivity of the mass surface to the shape evolution would be nevertheless observed at larger neutron number. The new result for $^{98}$Kr shows a slight, but definite, reduction of the $S_{2n}$ slope. 

In order to connect this trend to the underlying picture of intrinsic shapes, we resort to energy density functional (EDF) theory.
In Fig.~\ref{fig:DFT} (upper) the total energy is represented as a function of intrinsic quadrupole deformation parameter $\beta_{20}$ from an axial calculation using the UNEDF0 energy density functional~\cite{UNEDF0}. The calculation is performed for the even-even zirconium, strontium and krypton isotopes with $58 \le N \le 62$. The total energy surfaces (TES) for the same neutron number are overlapped, with the mean-field energy of the spherical calculation subtracted to obtain relative depths of the energy minima. Although the intrinsic nuclear shape is not a nuclear observable, within a certain model it offers a basis to link the trends of observables to predicted changes in nuclear structure. This is nicely illustrated in Fig.~\ref{fig:DFT} (upper), where for zirconium isotopes (in red) a spherical shape predominates for $N < 60$, but a strongly deformed and deep prolate minimum suddenly develops at $N = 60$ and beyond, driving the ground-state shape transition. The strontium isotopes (in orange) have a very similar behavior, which is also qualitatively consistent with the experimental data, whereas the krypton isotopes (in blue) show a markedly different evolution: a rather soft spherical-oblate minimum is present at $N= 58$, while at $N = 60$ and beyond, one oblate and two prolate minima develop, all of much more comparable depth. Although the prolate minimum at large deformation also appears, it is significantly less bound than in the case of strontium and zirconium, which confirms the more significant energy cost in krypton of promoting protons to the $g_{9/2}$ orbital.  

To illustrate the model dependence of the intrinsic shape landscape, we present in Fig.~\ref{fig:DFT} (lower) the equivalent TES plots for calculations using the Gogny D1S interaction. This interaction is interesting because it has been used extensively for calculations in the $A \approx 100$ region that includes the triaxial degree of freedom and beyond-mean-field (BMF) shape mixing \cite{Rodriguez-Guzman10,Rodriguez-Guzman10b,Delaroche10,Rodriguez2014}. It has also been used to calculate the energies of excited nuclear states in recent experimental work in the region \cite{Flavigny,clement,Dudouet}. 

First of all, one notices in Fig.~\ref{fig:DFT} (lower) that, unlike the UNEDF0 interaction, for D1S the deformed minima become increasingly deeper as one moves away from the zirconium isotopic chain. The reason for this behavior is that the $Z = 40$ shell closure is significantly larger for D1S (approximately 2.9~MeV, comparable to the $Z = 50$ gap) than for UNEDF0 (only 1.1~MeV).  Second, one observes the same predominance of the oblate minimum before $N = 60$. The prolate minimum significantly gains in binding energy from $N \ge 60$ and for zirconium and strontium, it becomes the most bound shape of the TES. For krypton, however, the oblate shape remains more bound until $N = 62$, although the difference to the prolate minimum diminishes.

Triaxial BMF calculations with interactions from the Gogny family have been performed in several previous works. This includes a global study of the even-even isotopes across the nuclear chart with a five-dimensional collective Hamiltonian (5DCH)~\cite{Delaroche10} and a detailed study of the even-even krypton isotopes using the Symmetry Conserving Configuration Mixing (SCCM) method~\cite{Rodriguez2014}. 
We have extended the SCCM calculations of~\cite{Rodriguez2014} to $^{100}$Kr, as illustrated in 
Fig.~\ref{TES-SCCM}, where the resulting triaxial deformation landscape obtained for the even-even krypton isotopes in this approach is shown. 
The top row presents the TES between $N = 50$ and $N = 64$. One can see that closer to $N = 50$ the TES has a weakly deformed triaxial minimum which evolves towards an oblate minimum at $N = 58$. From $N = 60$ a prolate minimum also emerges, becoming more pronounced at $N = 62$. This means that around $N = 60$ the axial picture from Fig.~\ref{fig:DFT} (lower) remains largely unchanged once the triaxial degree of freedom is open. At $N = 64$, however, a second, more deformed and prolate-like triaxial minimum develops near the axial one. 

The SCCM method allows to go beyond the mean-field picture explored so far, on the one hand restoring the angular momentum of the wave function and on the other hand allowing it to mix different shapes on the TES. The amplitudes of the different shapes in the final many-body solution define the so-called collective wave function. In rows two, three and four of Fig.~\ref{TES-SCCM}, this function is represented on the same triaxial deformation space for the first three $0^+$ states of the calculated krypton isotopes. One can observe that in the $0_1^+$ ground states up to $N=62$, the collective amplitude shows a predominance of the main mean-field minimum, triaxial up to $N = 56$ and smoothly transitioning to oblate-like in $N =60,62$. The deformed prolate minimum is present in the third $0^+$ state for $N = 58,60$ and in the second one for $N = 62$. Finally, for $N = 64$ the first two $0^+$ states completely change composition and become a mixture of the oblate-like and prolate-like triaxial shapes. 

The SCCM calculations thus predict that around $N = 60$ the ground state of the krypton nuclei remains oblate-like, with a sudden transition to mixed triaxial at $N = 64$. Unlike for the isotopic chains above, the ground state does not transition to prolate-like at $N = 60$.

\begin{figure}
  \centering 
\includegraphics[width = 0.975\linewidth]{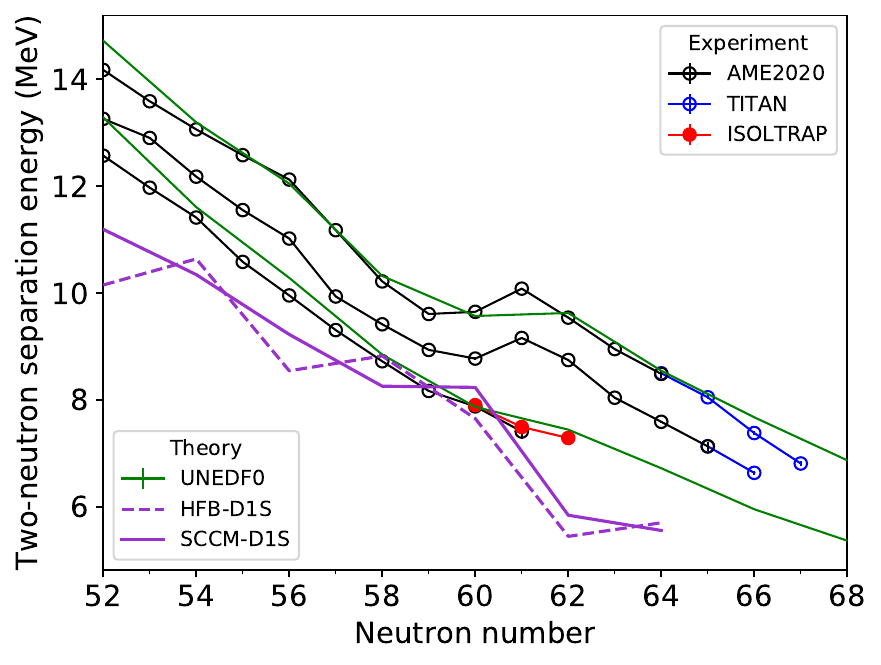}
\includegraphics[width = 0.975\linewidth]{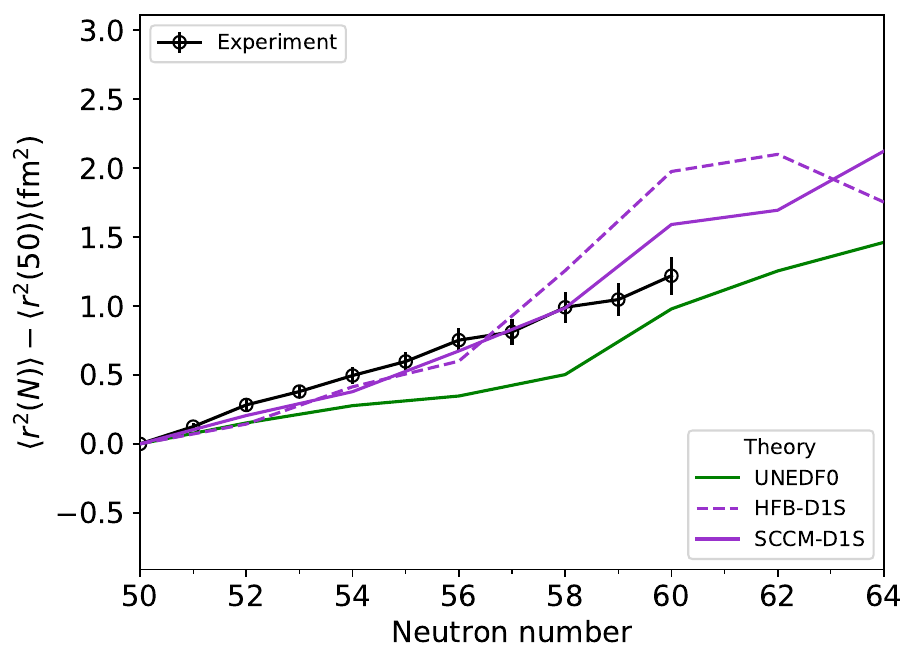}
 \caption{ Experimental two-neutron separation energies of krypton, rubidium and strontium isotopes (top panel) and differences in mean-square charged radii of krypton isotopes \cite{Keim} (bottom panel) represented as a function of neutron number and compared to calculations (of only even-even nuclides) using the UNEDF0 functional \cite{UNEDF0} (green) and the Gogny-D1S interaction (violet), the latter either in a triaxial HFB calculation (dashed line), or using the SCCM approach (continuous line).
 }
\label{S2n-rad-exp-th}
\end{figure}

Finally, we show the predictions of these various methods compared to the experimental observables in Fig.~\ref{S2n-rad-exp-th}.    
The UNEDF0 functional is shown in green and the Gogny D1S interaction, in violet (only even-even nuclei are calculated). The upper panel shows the two-neutron separation energies $S_{2n}$ for the krypton, rubidium and strontium isotopes, with the new experimental data from this work highlighted in red. The lower panel shows the difference in mean-square charge radii to $N = 50$, $\delta \langle r^2\rangle_{N,50}$. The UNEDF0 $S_{2n}$ values agree well with the experimental trends around $N = 60$, describing the change in behavior from strontium to krypton. In the latter, the slight change of slope at $N = 60$ is also reproduced, but as one can deduce from Fig.~\ref{fig:DFT}, this trend is due to the evolution of the binding energy of the oblate configuration and not due to a transition to a prolate one. The calculations of the charge radii show a sudden increase between $N = 58$ and $N = 60$, the consequence of passing from a soft oblate minimum to a rigid and slightly more deformed one. This trend is not observed in the experimental data, however.

The calculations using the Gogny D1S interaction
agree less with the experimental
data, but do allow observing the relative effect of taking into account beyond-mean-field shape mixing by the SCCM method. 
First, the $S_{2n}$ exhibits a systematic drift toward lower values compared to the experimental data. This discrepancy arises from the underestimation of binding energy by the D1S parametrization in neutron-rich nuclei~\cite{chappert2008}. More recent Gogny parametrizations, such as Gogny D1M~\cite{D1M}, correct this effect. However, deformation properties and beyond-mean-field correlations remain largely consistent across the most widely used Gogny energy density functionals~\cite{Robledo2011,Rodriguez2015}. Therefore, apart from the overall shift in $S_{2n}$, we do not expect significant differences when using alternative parametrizations.
For both the $S_{2n}$ and the $\delta \langle r^2\rangle_{N,50}$ values, the SCCM results (corresponding to the $0_1^+$ states in Fig.~\ref{TES-SCCM}) show a smoother evolution with neutron number than the static mean-field solutions (which correspond in the same figure to the most bound minimum on the TES). Both types of calculations show a significant change of structure at $N = 60$, where the oblate minimum passes from soft, to rigid and more deformed, driving a transition of the collective amplitudes from triaxial to strongly oblate. The predicted flattening of $S_{2n}$ due to this change of structure and the corresponding increase in $\delta \langle r^2\rangle_{N,50}$ are not confirmed by the existing experimental data. The fall-off of $S_{2n}$ at $N = 62$, due to the remarkable binding of $N = 60$ isotope, makes it difficult to observe the expected effect of the second change of structure at $N = 64$. However, no major change in the trend of charge radii is predicted. 

Interestingly, despite differences characterizing the deformation landscape in the $A \approx 100$ region, both the UNEDF0 and the D1S interactions predict that in the krypton isotopic chain the trends of the ground-state properties at least up to $A = 98$, are driven by the evolution of the oblate configuration, at odds with the higher $Z$ isotopic chains.  
So if the krypton isotopes are not a ``boundary'' as such, they would appear to reflect a transitional feature on the nuclear landscape.   



\section{Conclusion}


Masses of $^{96,97,98}$Kr were measured with the ISOLTRAP experiment at ISOLDE, $^{98}$Kr for the first time. This result extends the mass surface along what was proposed to be the lower-$Z$ boundary of the $A~=~100$ region of deformation. The resulting trend of two-neutron separation energies suggests a gradual onset of collectivity beyond $N = 60$, consistent with the recent spectroscopy of $^{98,100}$Kr at RIBF.  Axial calculations performed with the UNEDF0 functional, as well as  triaxial and beyond-mean-field calculations performed with the Gogny D1S interaction, predict that the structure of krypton, unlike the isotopic chains with larger proton number, is driven by the evolution of the oblate configuration. With this intrinsic picture, the UNEDF0 functional gives a good description of the measured $S_{2n}$ trend up to $N = 62$. A transition to a mixed, oblate-prolate structure is predicted by the beyond-mean-field calculations for $^{100}$Kr. 

\section{Acknowledgements}
\begin{acknowledgments}
We thank M. Breitenfeldt, S. George and A. de Roubin for help with the experiment.
We thank the ISOLDE technical group and the ISOLDE Collaboration for their assistance. 
We acknowledge support from the Max Planck Society, the German Federal Ministry of Education and Research (BMBF) (Contracts No.~05P12HGCI1, 05P15ODCIA, 05P15HGCIA, 05P18HGCIA, 05P21HGCI1 and 05P18RDFN1), the Deutsche Forschungsgemeinschaft (DFG, German Research Foundation) -- Project-ID 279384907 -- SFB 1245, the French IN2P3 
and the European Union’s Horizon 2020 research and innovation programme (Grant No. 654002 "ENSAR" and 682841 "ASTRUm"). J.K. acknowledges a Wolfgang Gentner Ph.D. scholarship (Grant No. 05E12CHA).  T.R.R. acknowledges support from the Spanish MICINN (Grant No. PID2021-127890NB-I00) and GSI-Darmstadt computing facilities.
The experiment was conducted by D.L., V.M., M.M., N.A.A., D.A., K.B., 
A.H., W.-J.H., J.K., I.K., Yu.A.L., D.N., M.R.,
T.S., A.W., F.W., and R.W.
Resources and supervision were provided by K.B. D.L. and L.S.  
Data analysis was performed by I.K., V.M., M.M., and L.N. 
AME calculations were performed by W.-J.H.
The SCCM calculations were performed by T.R.R.
The manuscript was prepared by D.L., V.M., M.M., and L.N.
All authors contributed to the editing of the manuscript. 
\end{acknowledgments}

\section{Data Availability}
The data that support the findings of this article are openly available:~  
https://zenodo.org/records/16778286

\bibliography{biblio}

\end{document}